\documentclass[journal,review,accept,oneauthors,pdftex,10pt,a4paper]{mdpi} 

\firstpage{1} 
\articlenumber{x}
\doinum{10.3390/------}
\pubvolume{18}
\pubyear{2016}
\copyrightyear{2016}
\externaleditor{Academic Editor: Takuya Yamano}
\history{Received:  3 May 2016; Accepted:  18 June 2016; Published: date}

\usepackage{soul}

 \theoremstyle{mdpi}
 \newcounter{thm}
 \newcounter{ex}
 \newcounter{re}
 \theoremstyle{mdpidefinition}

\newcommand{\x}{{\vec{x}}}
\newcommand{\y}{{\vec{y}}}
\newcommand{\z}{\vec{z}}
\newcommand{\bX}{{\vec{X}}}
\newcommand{\bY}{{\vec{Y}}}
\newcommand{\Y}{{\vec{Y}}}
\newcommand{\A}{{\mathsf{A}}}
\newcommand{\B}{{\mathsf{B}}}
\newcommand{\bC}{{\mathsf{C}}}
\newcommand{\DD}{{\mathsf{D}}}
\newcommand{\E}{{\sf{E}}}
\newcommand{\F}{{\mathsf{F}}}
\newcommand{\HH}{{\sf{H}}}
\newcommand{\bG}{{\mathsf{G}}}

\newcommand{\K}{{\sf{K}}}
\newcommand{\Q}{{\mathsf{Q}}}
\newcommand{\R}{{\mathsf{R}}}
\newcommand{\SSS}{{\mathsf{S}}}
\newcommand{\T}{{\mathsf{T}}}

\newcommand{\PP}{{\sf{P}}}
\newcommand{\UU}{{\sf{U}}}
\newcommand{\V}{{\sf{V}}}
\newcommand{\W}{{\sf{W}}}

\newcommand{\ZZ}{{\vec{Z}}}

\newcommand{\Tr}{{\rm Tr}}
\newcommand{\CXX}{{\sf{C}}_{\rm{XX}}}
\newcommand{\CXY}{{\sf{C}}_{\rm{XY}}}
\newcommand{\CYY}{{\sf{C}}_{\rm{YY}}}

\def\expec#1{\langle#1\rangle}

\graphicspath{{./figures/}}

\Title{Generalisations of Fisher Matrices}

\Author{Alan Heavens}
\AuthorNames{Alan Heavens}

\address[1]{%
Imperial Centre for Inference and Cosmology (ICIC), Imperial College London, Blackett Laboratory, \mbox{Prince Consort Road,} London SW7 2AZ, UK; {a.heavens@imperial.ac.uk}}

\corres{Correspondence: }

\abstract{
Fisher matrices play an important role in experimental design and in data analysis.  Their primary role is to make predictions for the inference of model parameters---both their errors and covariances.  In this short review, I outline a number of extensions to the simple Fisher matrix formalism, covering a number of recent developments in the field.  These are: (a) situations where the data (in the form of ($x,y$) pairs) have errors in both $x$ and $y$; (b) modifications to parameter inference in the presence of systematic errors, or through fixing the values of some model parameters; (c) Derivative Approximation for LIkelihoods (DALI) - 
 higher-order expansions of the likelihood surface, going beyond the Gaussian shape approximation; (d) extensions of the Fisher-like formalism, to treat model selection problems with Bayesian evidence.}

\keyword{fisher matrices; statistics; experimental design}

\begin{document}





\section{Introduction}

Fisher information matrices are widely used for making predictions for the errors and covariances of parameter estimates.  They characterise the expected shape of the likelihood surface in parameter space, subject to an assumption that the likelihood surface is a multivariate Gaussian when viewed as a function of the model parameters.  Diagonal terms are the inverse variances of the parameters, conditional on all others being known, and non-zero off-diagonal terms indicate correlations between inferred parameters.  Diagonal terms of the inverse Fisher matrix yield the variances of parameters when all others are marginalised over. The Cram\' er--Rao inequality shows that the variances deduced from the Fisher matrix are lower bounds.

Fisher matrices have been extensively used in cosmology, where future experiments have been designed in order to deduce as precisely as possible the parameters of the standard cosmological model, so-called $\Lambda$CDM (Cold Dark Matter, with a cosmological constant $\Lambda$), and are routinely used to give ``figures-of-merit'' \cite{DETF} for the power of each experiment.  Normally, these studies are standard applications of Fisher matrix theory, often simplified by an approximation (which is very good for observations of the Early Universe) that the data are Gaussian-distributed.

In this article, I review a number of generalisations of the Fisher matrix approach.  In {Section} {\ref{GRF}} the derivation of the Fisher matrix for Gaussian data is sketched out; in Section {\ref{FisherXY}} we consider Fisher matrices for data pairs that have errors in both $x$ and $y$; in Section \ref{Syst} we show how Fisher matrices may be used to estimate biases when some parameters are fixed at incorrect values; in Section \ref{DALI} we explore better approximations for the likelihood surface (``DALI''), from expansions to higher order in derivatives, and in Section \ref{Bayes} we generalise the use of Gaussian likelihood surfaces to model selection and Bayesian evidence.

\section{Gaussian Fields}\label{GRF}

In cosmology, one is very often dealing with Gaussian random fields, which are characterised statistically entirely by their mean and covariance.
A pedagogical derivation for the Fisher matrix when the data $\y$ are Gaussian appears in \cite{TTH}.  The negative log-likelihood ${\cal L}\equiv -\ln L$ is
\begin{equation}\label{GaussianEq}
2{\cal L}(\vec\theta) = \ln\det 2\pi\bC + (\vec{y}-\vec{\mu})^T
\bC^{-1}(\vec{y}-\vec{\mu}),
\end{equation}
where in general both the mean vector $\vec{\mu}$ and the covariance
matrix $\bC =\expec{(\vec{y}-\vec{\mu})(\vec{y}-\vec{\mu})^T}$
depend on the model parameters $\vec\theta$.  If $\y$ represents 1-point statistics, such as Fourier coefficients, then typically $\vec\mu=\vec 0$, and all the parameter dependence is in $\bC$.  If $\y$ represents 2-point statistics, then for Gaussian fields they have only approximately a Gaussian distribution, and the analysis is only approximately correct.  In this case, the covariance matrix has some parameter dependence through the 4-point function, which for Gaussian fields can be written as products of the 2-point function.

The Gaussian assumption is widely applicable in cosmology, since the quantum fluctuations that are thought to give rise to the density and radiation fields should ensure this, and limits on departures from gaussianity are very tight \cite{Planck2015NG}. Defining the data matrix $\DD\equiv (\vec{y}-\vec{\mu})(\vec{y}-\vec{\mu})^T$ 
and using the matrix identity for positive definite square matrices $\ln\det \bC = \Tr\ln
\bC$, where Tr indicates trace, we can re-write (\ref{GaussianEq})
as
\begin{equation}\label{GaussianEq2}
2{\cal L} = \Tr\left[\ln \bC + \bC^{-1}\DD\right].
\end{equation}

Using standard comma notation for partial derivatives, $Z_{,\alpha} = \partial Z/\partial\theta_\alpha$, and using the matrix identities
$(\bC^{-1}),_\alpha = -\bC^{-1}\bC,_\alpha \bC^{-1}$ and $(\ln \bC),_\alpha = \bC^{-1}
\bC,_\alpha$, we find after taking two derivatives and then the expectation value,
\begin{equation}
\F_{\alpha\beta} = \expec{{\cal L},_{\alpha\beta}} = \frac{1}{2}\Tr[\bC^{-1}\bC_{,\alpha}
\bC^{-1}\bC_{,\beta} + \bC^{-1}(\vec{\mu},_\alpha\vec{\mu},_\beta^T + \vec{\mu},_\beta\vec{\mu},_\alpha^T)],
\label{TTHF}
\end{equation}

The great advantage of the Fisher matrix approach is seen in this example: no data (real or simulated) are required to compute the expected log-likelihood surface, only the statistical properties of the data.  This can be a big advantage if simulation is computationally expensive. 

\section{Fisher Matrix with Errors in $x$ as Well as $y$}\label{FisherXY}

The previous section gives the standard analysis where only the covariance of the $y$ values is considered.  
Let us now consider the fairly general case where the data consists of data pairs $(X,Y)$, where we have errors in {\em both} $X$ and $Y$.  We can compute the Fisher matrix via the application of a Bayesian hierarchical model, provided that the errors in $X$ are small (this will be defined later). The full analysis is given in \cite{Heavens}.

We assume $\bX$ and $\bY$ are length $m$ and $n$ vectors (for data pairs, $m=n$, but in fact the analysis is more general), and have Gaussian errors, around true values $\x$, $\y$, with a covariance matrix $\bC$, which also allows correlations between $\bX$ and $\bY$.   $\x$ and $\y$ are
not observed,  being latent variables, and are essentially nuisance parameters.  In fact the $\y$ are not independent nuisance parameters as they are assumed to be related precisely to $\x$ through a deterministic theoretical model $\y=\vec{\mu}(\x)$ (however, a stochastic element could easily be included).
Given the observed data, $\bX,\bY$, we seek the posterior $p(\vec\theta|\bX,\bY)$.  With a uniform prior for $\vec\theta$, this is proportional to the likelihood $L = p(\bX,\bY|\vec\theta)$. 
We write this as the marginalised distribution over $\x$ and $\y$ as
\begin{eqnarray}
L &=& \int p(\bX,\bY,\x,\y|\vec\theta) \, d\x\, d\y 
= \int p(\bX,\bY|\x,\y,\vec\theta)p(\x,\y|\vec\theta)  \, d\x\, d\y\\\nonumber
&=&
\int p(\bX,\bY|\x,\y,\vec\theta)p(\y|\x,\vec\theta)p(\x|\vec\theta)  \, d\x\, d\y.
\end{eqnarray}

A deterministic $\y(\x)$ relation gives a delta function,
$p(\y|\x,\vec\theta)=\delta(\y-\vec\mu(\x))$, and assuming  a uniform prior for $\x$ (a more general prior is considered in \cite{Heavens}),
 integration over $\y$ gives
\begin{equation}
L \propto \int   p(\bX,\bY|\x,\vec\mu(\x),\vec\theta)  \, d\x.
\end{equation}

We now assume that the errors in $\bX$ are small, for which we require that we can truncate at the
linear term of the Taylor expansion of $\vec\mu$:
\begin{equation}
\vec\mu(\x)=\vec\mu(\bX)+\T(\bX)\,(\x-\bX),
\end{equation}
where
\begin{equation}
\T_{ij} \equiv \left.\frac{\partial \mu_i}{\partial x_j}\right |_{\x=\bX}.
\end{equation}

$\T$ is diagonal for data consisting of $\bX,\bY$ pairs.

We assume a multivariate Gaussian for $p(\bX,\Y|\x,\y)$ (independent of $\vec\theta$), and write the covariance matrix of the data in block form as
\begin{equation}
\bC = \bordermatrix{~ & \bX & \Y \cr
                  \bX & \CXX & \CXY \cr
                  \Y & \CXY^T & \CYY \cr}.
\end{equation}

Note that $\bC_{\rm{XY}}\equiv \expec{(\vec{X}-{\bar{\vec{X}}})(\vec{Y}-\bar{\vec{Y}})^T}$ is not symmetrical, nor invertible or even square in general; although $\bC_{\rm{XX}}$ and $\bC_{\rm{YY}}$ are.    The covariance matrix may include a number of elements, such as intrinsic scatter and measurement noise, with individual covariance matrices adding to give the final $\bC$.  We also assume that the function $\vec\mu(\x)$ is linear
across the width of the Gaussian error distribution of $\x$, in which case the likelihood may be integrated analytically, as follows.  We write
\begin{equation}
L \propto \int
\frac{1}{\sqrt{\det\bC}}\exp\left(-\frac{Q}{2}\right)\,d\x \label{eqn:likelihood_gaussian}
\end{equation}
where $Q \equiv (\ZZ-\z)^T\bC^{-1}(\ZZ-\z)$, and $\z$ and $\ZZ$ are $m+n$-dimensional vectors:  $z_i=x_i$ and $Z_i=X_i$  for $i\le m$, $Z_{m+j}=Y_j$ and $z_{m+j}=\mu_j(\bX)+[\T(\bX)(\x-\bX)]_j$.
The inverse of $\bC$ in block form is
\begin{equation}
\bC^{-1} = \left(\begin{matrix}
\bG & -\HH \cr
-\HH^T & \E
\end{matrix} \right)
\end{equation}
where
\begin{eqnarray}
\bG &=& \bC_{\rm{XX}}^{-1}+  \bC_{\rm{XX}}^{-1}\bC_{\rm{XY}}\E \bC_{\rm{XY}}^T \bC_{\rm{XX}}^{-1}\cr
\HH &=&  \bC_{\rm{XX}}^{-1}\bC_{\rm{XY}}\E\cr
\E &=& (\bC_{\rm{YY}}-\bC_{\rm{XY}}^T\bC_{\rm{XX}}^{-1}\bC_{\rm{XY}})^{-1}.
\end{eqnarray}

Defining $\tilde \x \equiv \bX-\x$, and $\tilde \Y\equiv \Y-\vec\mu(\bX)$, we find that
$Q$ has the  quadratic form
\begin{equation}
Q = \tilde \x^T \A \tilde \x - \B^T\tilde \x  - \tilde \x^T\B+ Q', \label{eqn:qdefinition}
\end{equation}
where
\begin{eqnarray}
\A &=& \bG + \T^T\E\T - \HH\T - \T^T\HH^T\nonumber\\
\B &=& (\HH-\T^T\E)\tilde \Y \equiv \PP\tilde \Y\nonumber\\
Q' &=&  \tilde \Y^T\E \tilde \Y.
\end{eqnarray}

With the definition of $Q$ in Equation ~(\ref{eqn:qdefinition}), the Gaussian integral of Equation ~(\ref{eqn:likelihood_gaussian}) can be performed, using
\begin{equation}
\int \exp\left({-\frac{1}{2}\tilde \x^T\A\tilde \x +\B^T\tilde
  \x}\right) d\tilde \x =
\frac{(2\pi)^{n/2}}{\sqrt{\det{A}}}\exp\left({\frac{1}{2}\B^T\A^{-1}\B}\right),
\end{equation}
\noindent and noting that $Q'$ is independent of $\tilde x$.
The likelihood then simplifies to
\begin{equation}
L \propto \frac{1}{\sqrt{\det\R}}\exp\left(-\frac{1}{2} \tilde \Y^T \R^{-1} \tilde \Y\right),
\end{equation}
where the inverse of the marginal covariance matrix of $\tilde \Y$ is $\R^{-1} = \E-\PP^T\A^{-1}\PP$. This is obtained using the
Woodbury formula \citep{1950woodbury} $(\K + \UU\W\V)^{-1} = \K^{-1}- \K^{-1}\UU(\W^{-1}+\V\K^{-1}\UU)^{-1}V\K^{-1}$, giving
\begin{equation}
\R = \bC_{\rm{YY}}-\bC_{\rm{XY}}^T\T^T -\T\bC_{\rm{XY}}+\T\bC_{\rm{XX}}\T^T,
\label{Rmatrix}
\end{equation}

This is a key result.  We see that this looks just like a standard Gaussian (in terms of data) likelihood, but with the covariance matrix $\bC$ ($\CYY$ in our current notation) replaced by $\R$.  Hence to compute the Fisher matrix, we can use the standard formula found in Equation ~(\ref{TTHF}) and \mbox{Equation  (15)} of \cite{TTH}, and simply replace $\bC$ by $\R$:
\begin{equation}\label{Fnew}
\F_{\alpha\beta} = \frac{1}{2}{\rm{Tr}}\left[\R^{-1}\R_{,\alpha}\R^{-1}\R_{,\beta} + \R^{-1}(\vec\mu_{,\alpha}\vec\mu_{,\beta}^T+\vec\mu_{,\beta}\vec\mu_{,\alpha}^T)\right].
\end{equation}

Note that $\R$ depends not only on the standard covariance, but also on the covariance in the
independent variable, $\bC_{\rm{XX}}$, the meta-covariance, $\bC_{\rm{XY}}$, and
the first partial derivatives of the model function $\vec\mu$.  In the case of uncorrelated data pairs, the result reduces to that found in \cite{March}.  For the simple case of no correlations between $\bX$ and $\Y$ values $\R=\CYY+\T^T\CXX\T$, and with diagonal covariance matrices $\bC_{\rm{YY}}$ and  $\bC_{\rm{XX}}$ we recover the propagation of error result that the variance of $f\equiv Y-\mu(X)$ for each data point is effectively
\begin{equation}
\sigma_f^2 = \sigma_{\rm Y}^2 + \mu'(X)^2\,\sigma_{\rm X}^2,
\end{equation}
where $\mu' = \partial\mu/\partial x$ and $\bC$ can be replaced in the standard Fisher expression, Equation  (\ref{TTHF}), by a diagonal $ n \times n$ matrix with these enhanced entries. 

\subsection*{Generalising Still Further}

The analysis above is applicable not just to the simple case of data with errors in $x$ as well as $y$, but to any system where the `data' $\y$ depend (in a locally linear way) on {\em any} parameters $\x$ that have some error associated with them.  

\section{Systematic Errors, or Errors from Simplified Nested Models}\label{Syst}

\begin{figure}[H]
\centering
\includegraphics[width=8cm]{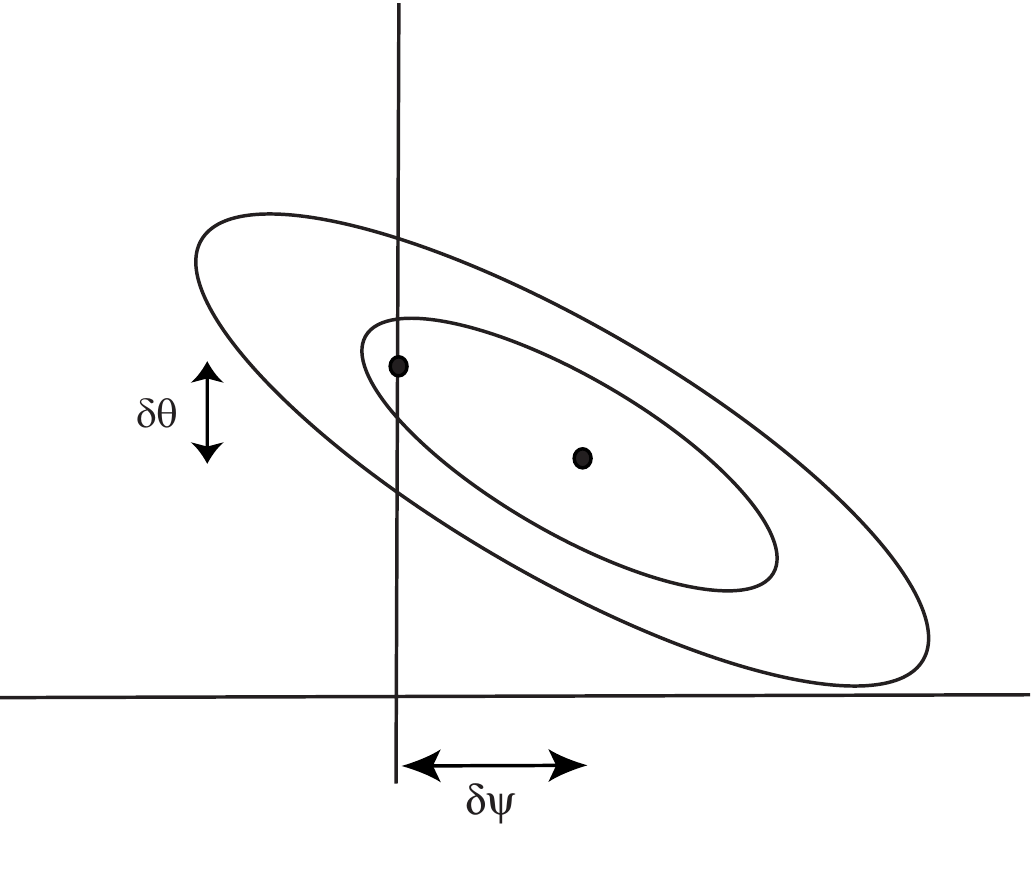}
\caption{Illustration of the shift in the maximum likelihood value of a parameter, if another parameter is set at a specific value (e.g., a calibration parameter, or a nested model which does not allow a parameter to vary from some fixed value.  The Fisher matrix may be used to determine the shift. Reproduced from Figure 1 of {\em ``On model selection forecasting, Dark Energy and modified gravity''} published in Mon. Not. Roy. Astron. Soc. \cite{HKV}.  }
\label{Bias}
\end{figure}

The Fisher matrix can also be useful to determine the errors in parameter inference that arise if one parameter is fixed at an erroneous value.  This could arise in a number of contexts, such as a nuisance parameter (e.g., a calibration setting) being fixed at an incorrect value, or when considering nested models.  An example of the latter would be cosmological models where the Universe is assumed to be flat.  This is an example of a nested model, being a subset of a more general model, but with the curvature parameter (usually given the symbol $\Omega_k$) set to zero.  In these cases, the maximum likelihood values of all the other parameters are, in general, shifted from their maximum likelihood values in the more general model.  See Figure \ref{Bias} for an illustration of this in two dimensions. With the usual Fisher assumption that the likelihood surface is a Gaussian function of the parameters, these shifts can be computed using the Fisher matrix.  

We consider two models, $M$, which has more ($n'+p$) parameters than a simpler nested model $M'$, which has $n'$. The extra parameters are designated $\psi_\zeta$, and these are fixed in $M'$ at values that are $\delta\psi_\zeta$ from their maximum likelihood values in $M$.  In this case, the maximum likelihood values of all other parameters of $M'$, $\vec\theta$, are systematically shifted by \cite{Taylor2007, HKV}
\begin{equation}
\delta\theta_\alpha =
-(\F'^{-1})_{\alpha\beta}\bG_{\beta\zeta}\delta\psi_\zeta \qquad
\alpha,\beta=1\ldots n', \zeta=1\ldots p \label{offset}
\end{equation} where
\begin{equation}
\bG_{\beta\zeta}=\frac{1}{2}{\rm
Tr}\left[\bC^{-1}\bC_{,\beta}\bC^{-1}\bC_{,\zeta}+\bC^{-1}(\vec\mu_{,\zeta}\vec\mu^T_{,\beta}+\vec\mu_{,\beta}\vec\mu^T_{,\zeta})\right],
\label{eqG}
\end{equation}
which we recognise as a subset of the Fisher matrix.

\section{Beyond the Gaussian Approximation---DALI}\label{DALI}

The Fisher matrix approach assumes that the likelihood surface is a multivariate Gaussian, which will be asymptotically true near the peak, but may not be a good approximation over the range of parameter values of interest.  A generalisation of the Fisher matrix is DALI, Derivative Approximation for LIkelihoods \cite{SQA}, which expands the likelihood surface to include higher-order derivatives than the second.  This is a rather elegant expansion, in derivatives rather than parameters, that ensures that the approximate distribution is a genuine probability distribution---\textit{i.e.}, it is non-negative and normalisable, non-divergent and asymptotically approaches the true likelihood. 

The starting point is a Taylor expansion of the likelihood:
\begin{equation}
\begin{aligned}-\mathcal{L}\equiv\ln L\approx & \ln L_{0}+\frac{1}{2!}\F_{\alpha\beta}\;\Delta \theta_{\alpha}\Delta \theta_{\beta}
 +\frac{1}{3!}\SSS_{\alpha\beta\gamma}\;\Delta \theta_{\alpha}\Delta \theta_{\beta}\Delta \theta_{\gamma}\\
 & +\frac{{1}}{4!}\Q_{\alpha\beta\gamma\delta}\;\Delta \theta_{\alpha}\Delta \theta_{\beta}\Delta \theta_{\gamma}\Delta \theta_{\delta}+\ldots,
\end{aligned}
\end{equation}
where $L_0$ is a normalization constant and $\F_{\alpha\beta} =\mathcal{L}_{,\alpha\beta}$, $\SSS_{\alpha\beta\gamma} 
 =\mathcal{L}_{,\alpha\beta\gamma}$ and $\Q_{\alpha\beta\gamma\delta}  =\mathcal{L}_{,\alpha\beta\gamma\delta}$.

If the expansion is arranged  in order of derivatives, the expressions are normalisable and positive-definite. For example, to second order in the $\vec{\mu}$ derivatives, and assuming $\bC$ is independent of $\vec\theta$,  we have
\begin{equation}
\begin{aligned}L=L_0\exp\bigg[ & -\frac{1}{2}{\vec\mu}_{,\alpha}^T\bC^{-1}{\vec\mu}_{,\beta}\Delta \theta_{\alpha}\Delta \theta_{\beta}  - \bigg(\frac{1}{2}{\vec\mu}_{,\alpha\beta}^T \bC^{-1}{\vec\mu}_{,\gamma}\Delta \theta_{\alpha}\Delta \theta_{\beta}\Delta \theta_{\gamma}\\
 & +\frac{1}{8}{\vec\mu}_{,\delta\gamma}^T \bC^{-1}{\vec\mu}_{,\beta\alpha}\Delta \theta_{\alpha}\Delta \theta_{\beta}\Delta \theta_{\gamma}\Delta \theta_{\delta}\bigg) +\,{\cal O}(3)\,\bigg].
 \end{aligned}
\label{eq:exp-1}\end{equation}

\begin{figure}[H]
\centering
\includegraphics[width=15cm, angle=0]{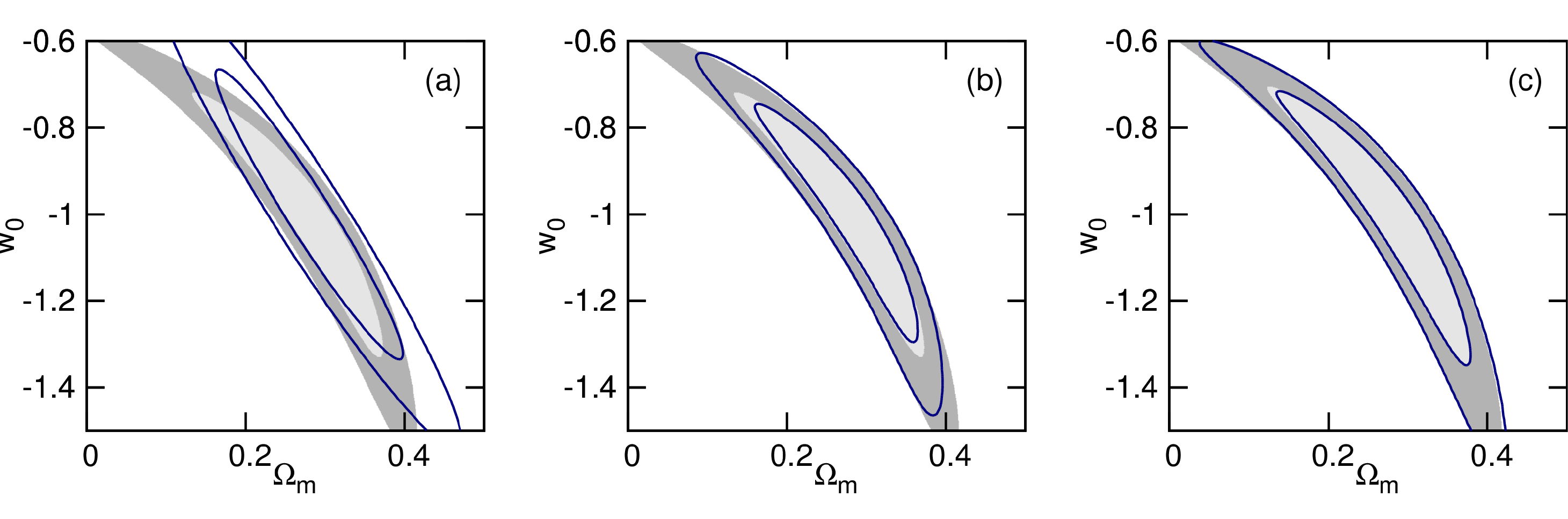}
\caption{The extended Fisher likelihood of DALI 
as applied to a sample of supernovae,  as a function of the matter density parameter of the Universe, $\Omega_m$, and the equation of state parameter $w_0 = p/\rho$ of dark energy.  From left-to-right this shows the Fisher approximation, the ``doublet-DALI'' expansion, and the ``triplet-DALI'' expansion, all compared with the likelihood evaluated on a grid. Reproduced from Figure 1 of {\em ``Breaking the Spell of Gaussianity: Forecasting with Higher Order Fisher Matrices''} published in Mon. Not. Roy. Astron. Soc. \cite{SQA}.}
\label{figDALI}
\end{figure}

This is apparently true at every order (see \cite{SQA} for the third-order expansion, and \cite{Sellentin} for the case where the parameter dependence is in $\bC$). Figure \ref{figDALI} shows the improvement in the expected likelihood surfaces for a supernova cosmology experiment.

\section{The Expected Bayesian Evidence---Generalising Fisher Matrices to Model Selection}\label{Bayes}

At the root of the Fisher matrix formalism is the Laplace approximation, \textit{i.e.}, the assumption that the likelihood surface is a multivariate Gaussian when viewed as a function of the model parameters.  We can generalise this to the higher-level question of {\em model selection}, where we compute the posterior probabilities of different models, given the data collected, but regardless of the model parameters $\vec\theta$. The ratio of these probabilities is the ratio of the prior model probabilities, multiplied by the ``Bayes factor'', which is the ratio of the marginal likelihoods (or Bayesian evidence) of the models, where the evidence for a model $M$ is
\begin{equation}
p(\vec y|M) = \int p(\vec y|\vec\theta, M)p(\vec\theta | M)\,d\vec\theta.
\end{equation}

With the Laplace approximation for the first, likelihood term, and a uniform prior (which can be generalised to a Gaussian prior), we can compute the expected evidence (conditional on some fiducial set of parameters) by performing Gaussian integrals.  For nested models (with $n'$ and $n=n'+p$ parameters respectively), the considerations of Section \ref{Syst} on the locations of the peak likelihood is relevant, and the result depends on the shifts of the fiducial parameters away from the values that are fixed in the lower-dimensional model, $\delta\psi_\zeta$.  If we further approximate that the expected Bayes factor is the ratio of the expected evidences, then the expected Bayes factor is (see \cite{HKV} for details)
\begin{equation}
\langle B \rangle =
(2\pi)^{-p/2}\frac{\sqrt{\det{\F}}}{\sqrt{\det{\F'}}}\exp\left(-\frac{1}{2}\delta\theta_\alpha
\F_{\alpha\beta}\delta\theta_\beta\right)\prod_{q=1}^p\Delta\theta_{n'+q}
\label{Final}
\end{equation}
where $\Delta\theta_\alpha$ are the prior ranges of the additional $p$ parameters in the extended model, and the offsets $\delta\theta_\alpha$ are given by Equation (\ref{offset}).
Note that $\F$ is an $n \times n$ matrix, $\F'$ is $n'
\times n'$, and $\bG$ is an $n' \times p$ block of the full $n \times
n$ Fisher matrix $\F$, given by Equation (\ref{eqG}).  The expression we find is a specific example
of the Savage-Dickey density ratio \cite{Dickey}; here we explicitly use
the Laplace approximation to compute the offsets in the parameter
estimates which accompany the wrong choice of model.

\begin{figure}[H]
\centering
\includegraphics[width=7cm]{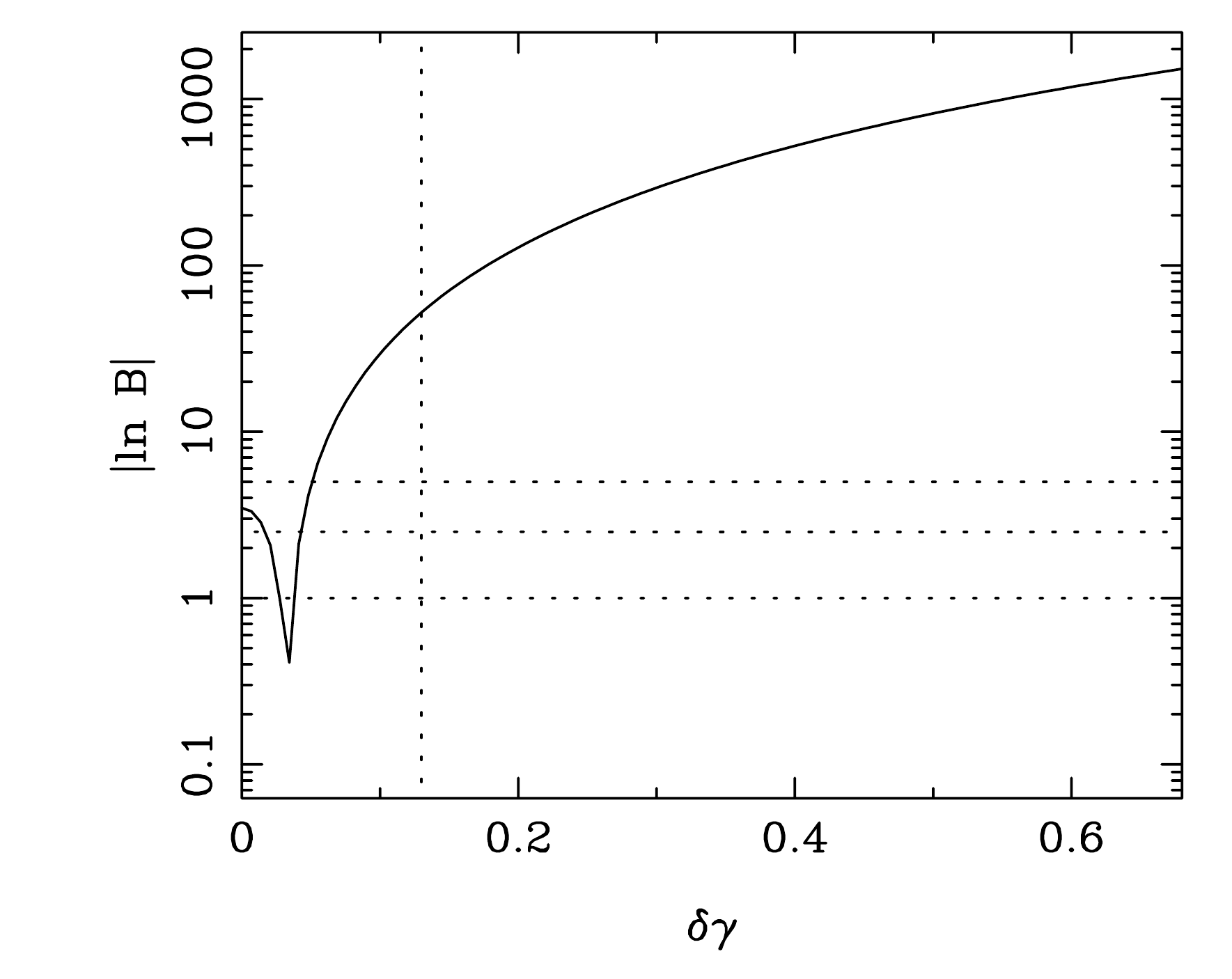}
\caption{The ratio of expected evidences $B$ for two cosmological models. One is based on Einstein Gravity; the other is a more general model where the growth rate of perturbations is allowed to be a free parameter, rather than fixed. The graph shows the ratio as a function of the true shift of the growth rate away from the General Relativity value, for weak lensing data expected from ESA's
 Euclid satellite.  If the growth rate is close to Einstein's (left of the figure; $\ln B>0$), Bayesian evidence is expected to favour Einstein gravity, whereas if the deviation is large enough (right of the cusp; $\ln B<0$), it favours the more complex model. Adapted from Figure 2 of {\em ``On model selection forecasting, Dark Energy and modified gravity''} published in Mon. Not. Roy. Astron. Soc. \cite{HKV}. }
\label{BayesFactor}
\end{figure}

Figure \ref{BayesFactor} shows the ratio of expected evidences, assuming the Laplace approximation (as the Fisher matrix does), for nested cosmological models.  Details are in the caption, but essentially one parameter is fixed in the simpler model, but allowed to vary in the more complex model. If the more complex model applies, then the data will favour the simpler model if the parameter is close to the fixed value.  This is shown in the figure by the cusp in the graph.  $\ln B$ is positive to the left of the cusp, and negative to the right.


\section{Discussion}

This article reviews some recent developments in Fisher matrix theory, which have been motivated by cosmology. The Fisher matrix for data consisting of pairs that have errors in both $x$ and $y$ is derived, as a specific example of a general result where the data can depend on arbitrary variables $x$ that may be measured with some error.  The Fisher matrix is shown to be able to determine biases in some parameters when others are set to fixed values (such as in nested models where the simpler model does not allow some parameters to vary).  DALI, which goes beyond the Laplace approximation by using higher-order derivatives, is found to allow much more accurate predictions for the expected shape of the likelihood surface.  Finally, the concept of expected probabilities in the Laplace approximation is generalised to model selection, by computing the expected \mbox{Bayesian evidence.}

%
%

%

\vspace{6pt} 

\acknowledgments
{The author thanks Elena Sellentin for useful discussions about DALI.}

\conflictofinterests{{The author declares no conflict of interest.}} 


\bibliographystyle{mdpi}

\newpage
\renewcommand\bibname{References}


\end{document}